\documentstyle[12pt,aasms4]{article}

\begin{document}
\title{The Fundamental Plane of Elliptical Galaxies in Compact Groups}

\author{Ignacio G. de la Rosa\altaffilmark{1}, 
Reinaldo R. de Carvalho\altaffilmark{2,}\altaffilmark{3}}
\and
\author{Stephen E. Zepf\altaffilmark{4}}

\altaffiltext{1}{Instituto de Astrof\'\i sica de Canarias, E-38200 La Laguna, Tenerife, Spain; irosa@ll.iac.es}
\altaffiltext{2}{Departamento de Astrofisica, Observat\'orio Nacional, CNPq, 20921-400 Rio de Janeiro, Brazil; reinaldo@voyager.on.br}
\altaffiltext{3}{Instituto Astron\^omico e Geof\'\i sico - USP, 04301-904 S\~ao Paulo, Brazil; reinaldo@iagusp.usp.br}
\altaffiltext{4}{Department of Astronomy, Yale University, P.O. Box 208101, New Haven, CT 06520-8101; zepf@astro.yale.edu}

\begin {abstract}
We present a study focusing on the nature of compact groups through the study of their elliptical galaxies. We determine velocity dispersions ($\sigma$) for 18 bright elliptical galaxies located in the core of Hickson compact groups and a control sample of 12 bright 
{\it bona fide\/} ellipticals located in the field or very loose groups. Several tests are carried out to avoid sources of systematic effects in $\sigma$ measurements. We use these velocity dispersions to compare the position of 11 compact group galaxies in the Fundamental Plane to that of a large and homogeneous sample of elliptical galaxies (Burstein et al. 1987). We find that little or no significant difference exists, as far as the Fundamental Plane is concerned, between ellipticals in compact groups and their counterparts in other environments.  
\end{abstract}


\section{Introduction}

In the last decade many independent studies of the global properties of elliptical galaxies
have established an observational basis for studying some of the underlying physical mechanisms of galaxy formation or at least one aspect of it ({\it e.g.,} Dressler et al. 1987;
Djorgovski \& Davis 1987; Faber et al. 1989; de Carvalho \& Djorgovski 1992; J\o rgensen, Franx, \& Kj\ae rgaard 1995a,b; Pahre 1998). In essence, the properties of elliptical galaxies occupy a planar surface within the 3-D space defined by the effective radius of the galaxy (R), the surface brightness within the aperture defined by this radius ($\mu$, or $I$ in linear flux units), and the central velocity dispersion, $\sigma_{\circ}$. This is the Fundamental Plane (FP), which represents a scaling relation between characteristic parameters describing the elliptical galaxies as a family. This relation has been used to estimate distance (Lynden-Bell et al. 1988), although its use as a tool to study the formation and evolution of elliptical galaxies has proved to be very effective as well.

Several authors have tried to measure systematic differences of the FP when defined using
elliptical galaxies populating different clustering scales ({\it e.g.,} de Carvalho \&
Djorgovski 1992; Guzm\'an et al. 1992). More recently, J\o rgensen, Franx, \& Kj\ae rgaard
(1996) have undertaken a photometric and spectroscopic survey on early-type galaxies in
order to define the FP for ellipticals in clusters reliably, establishing that
 $R\sim \sigma_{\circ}^{1.20} I^{-0.83}$. Pahre (1998) has
conducted a similar survey in the infrared and found that  $R\sim \sigma_{\circ}^{1.53}  I^{-0.79}$. Zepf \& Whitmore (1993, hereafter ZW93) were the first to measure the FP for ellipticals in compact groups (CGs), and found tentative evidence that it differs from that for  clusters because these galaxies in CGs have lower velocity dispersions. They interpreted this difference as an indication that the environment of the CGs has changed the structure and dynamics of ellipticals inhabiting them. Given the importance and implications of such a finding, we re-examine this topic with a better and more homogeneous dataset. 

The paper is organized as follows: \S  2 presents the observations and their basic reduction; \S  3 discusses the methods employed to measure velocity dispersion and the pitfalls present in the measurement process; \S  4 presents the external comparison with other sources of the literature; \S  5 shows the comparison with the FP, and \S  6 contains a discussion and the main conclusions of this work.

\section{Observational Data}
\subsection{Observations}

Our sample consists of 22 bright elliptical galaxies located in the cores of Hickson compact groups 
(hereafter the HCGs; Hickson 1982) and 12 bright {\it bona fide\/} E/S0s located in the field or very loose groups (hereafter the NGCs). 
Data were obtained during a four night run (1994 January 9--12)
 at the KPNO 2.1 m 
telescope with the GoldCam CCD spectrometer. A 600 line/mm grating was used to yield a 
$\sim$ 1.25 \AA\ /pixel dispersion over the wavelength range 3500  to 7000 \AA. The long-wavelength range and the intermediate spectral resolution (4.25 \AA\ FWHM) allows us to measure both the velocity dispersion with an ample spectral range and the whole set of 
spectral indices defined between the 4000 \AA\ break, $D_{\rm 4000}$, and the TiO band ($\sim$ 6200 \AA).

Table 1 shows the list of observed objects (column 1) followed by their 
morphological types (column 2),
taken from Hickson (1994; HCG galaxies) and from de Vaucouleurs et al. (1991;
NGC galaxies). Exposures (in seconds) are shown both for each night (columns 3 to 6) and for the total run (column 7). The apparent magnitude of the galaxies (column 8) was 
taken from Hickson (1994; HCG galaxies) and, for the NGC galaxies,
 from Burstein et al (1987;
hereafter B87). 
Signal-to-Noise ratio (S/N)---column 9---is calculated from Poisson statistics, using the gain (2.8 e$^-$/ADU), readout-noise (8.5 e$^-$), number of combined spectra, and sky value. The listed S/N values are
 the result of the average along a 1000 \AA\ interval, centered on 
the Mg lines at 5175 \AA\ and given as S/N per \AA.

The observations are long-slit spectra of galaxies and calibration stars with a fixed slit-width of 2 arcsec and a 0.78 arcsec/pix scale at the focal plane. The choice of a proper aperture size in the spatial direction results from a compromise among several constraints: highest S/N, smallest central aperture that is larger than the seeing limit, and need for
 minimal aperture correction. We must also ensure that, before their combination, the whole set of individual spectra of each object have  similar apertures, regardless their different exposure times. Seeing estimates, obtained from the FWHM of the spatial profile of the stellar spectra, are: 3.9 pix (day 01/09); 5.1 pix (01/10); 5.3 pix (01/11), and 5.2 pix (01/12). As a compromise, a constant aperture of 10 pixels was chosen for the whole set of observations. 

Also included in Table 1 (column 10) is a relevant parameter, used later (\S  4.1) for the aperture correction, the log ($r_{\rm  ap}/r_{\rm norm}$) (J\o rgensen et al. 1995b). Where $r_{\rm  ap}$ is the angular radius of an equivalent circular aperture:  
$r_{\rm ap} \sim 1.025 (xy/\pi)^{1/2}$,  $x$ and $y$ being
 the width and length of the aperture. Therefore, our fixed angular $r_{\rm ap}$ value is 2.284 arcsec, with $x = 2$ arcsec, $y = 7.8$ arcsec (10 pix $\times$ 0.78 arcsec/pix). The conversion from angular into metric radius is based on distances to the galaxies, derived from their radial velocities, published by Hickson (1994; HCG galaxies) and Burstein et al. (1987; NGC galaxies). On the other hand, $r_{\rm norm}$ is a fixed metric radius defined as $r_{\rm norm}$ = 0.595 $h^{-1}$ kpc, equivalent to $r_{\rm ap}$ = 1.7 arcsec at the distance of the Coma cluster.

\placetable{T1}

\subsection{Basic Reduction}

The spectra were reduced following standard procedures with the IRAF image processing tools, briefly described here. The bias subtraction was carried out using both the overscan region and the bias frames. Interpolations over non-linear pixels were applied after the construction of a bad-pixel mask. A group of some 80 partially bad columns was removed from the spectral region above 6600 \AA\ (rest frame). Exceptions are the two spectra with the largest redshift (HCG 28b and HCG 32a), where the damage included their H$\alpha$ line (6563 \AA). Dome flats were combined and normalized by fitting a function to remove large-scale variations of the lamp along the dispersion direction. No fringing pattern was evident. The sky background was removed by subtracting the low-order function which best fitted the signal, to both sides of the object, along the spatial direction. Immediately before the background subtraction, each sky spectrum is preserved for further S/N calculation.

The frames were cleaned for cosmic rays (CRs) through the following two steps. First, obvious CR candidates were removed from the background by direct image edition. Later, CRs inside the spectrum were carefully identified and  removed only when they showed both limited spatial extent and odd spectral positions.

The aperture extraction of the flux into a one-dimensional spectrum is made through the sum of all the sky-subtracted pixels, without using weighting. The use of variance weighting would distort the light profile, making the ``aperture correction'' meaningless. The IRAF task APALL uses a trace function, along the brightest point of the spatial profile, to center the apertures at each dispersion point of the image. This center does not always coincide with the center of the brightest pixel and, therefore, the aperture (10 pixels) does not generally contain an integral number of pixels. The signal from those fractional pixels is calculated as the contribution from the fractional area. Although dealing with fractional pixels is potentially dangerous in the presence of the steep radial gradients of elliptical galaxies, this is not expected in our case, due to the strong blurring of spatial information across individual pixels. Nevertheless, to test this effect, we have compared $\sigma$ measurements obtained from different 1-D spectra, which were extracted from the same 2-D spectrum, whose aperture differed by an amount much smaller than the size of the blurring profile. We found that an increase of 0.5 pixels in the aperture radius is typically accompanied by a decrease of less than 0.7 \% in the velocity dispersion.

The wavelength calibration is made with He--Ne--Ar lamp spectra taken 
immediately before or after the object spectra. The average rms scatter of the calibration is 0.08 \AA\ , equivalent to an uncertainty of 5 km s$^{-1}$ in the radial velocity. The spectra were logarithmically binned in wavelength.  

The instrumental resolution is measured by fitting Gaussians to the emission lines of the He--Ne--Ar lamp spectra and  shows a slight dependence with wavelength. The average $\sigma_{\rm instr}$ first increases by $\approx$ 0.2 \AA\ between 4500 and 5500 \AA\ and then decreases by $\approx$ 0.3 \AA\ between 5500 and 7000 \AA. The average value along the whole spectral interval is $\sigma_{\rm instr}$ = 1.75 \AA, equivalent to 105 km s$^{-1}$ at 5000 \AA. 

Individual exposures were added to create the combined final spectra. Extra care was taken to avoid an artificial broadening of the spectral lines due to poor matching of the wavelength scales of the individual spectra.

\section{Measurement of Velocity Dispersions}

Three methods exist to determine velocity dispersion ($\sigma$) from early-type galaxies using stellar templates (for a review see Rix \& White 1992). Two of these were used in the present work, with the aim of gaining internal consistency rather than of comparing their performances. First method (IRAF/FXCOR task), called CC for cross-correlation, determines $\sigma$ from the width of the cross-correlation peak between galaxy and template star (Tonry \& Davis 1979). The second method, called FF for Fourier Fitting, fits the Fourier transform of the galaxy spectrum with that of the template star convolved with the Gaussian broadening function (Franx, Illingworth, \& Heckman 1989). 

The first step for both procedures was to bring the spectra of the template stars to rest in the heliocentric system
and then measure radial velocities for the galaxies with respect to the templates (IRAF/FXCOR task). The combined spectra of galaxy were then reduced to the rest frame.  

\subsection{Cross-Correlation Method}

\subsubsection{The Calibration}

This method needs a calibration to translate its output, {\it i.e.,} the width ($\sigma_{\rm  peak}$) of the Gaussian curve fitting the cross-correlation peak, in terms of the velocity dispersion of the galaxy, $\sigma_{\rm gal}$. To construct the calibration curve, each stellar template is convolved with a set of Gaussian functions of varied $\sigma$, to simulate galaxies with different velocity dispersion. The cross-correlation between the simulated galaxies and the stellar template gives the FWHM$_{\rm peak}$ versus $\sigma_{\rm gal}$ calibration. In our case, this empirical calibration closely matches $\sigma_{\rm peak} = \sqrt {\sigma_{\rm gal}^2 + 2 \times \sigma_{\star}^2}$, with $\sigma_{\star}$ corresponding to $(2)^{-1/2} \times$ the width of the cross-correlation peak for the template--template correlations. The value of the $\sigma_{\star}$ shows a slight dependency with the spectral interval used in the cross-correlation. For the 4100--6300 \AA\  region, $\sigma_{\star} \approx$ 133 km s$^{-1}$.  

\subsubsection{The Peak Width}

As pointed out by Dalle Ore et al. (1991), one of the most delicate steps of the CC method is the isolation and measurement of the central peak in the cross-correlation function. This narrow peak, due to single lines, carries all the relevant information on velocity dispersions. However, the peak itself is superimposed upon an asymmetrical pedestal produced by blends and residual trends that remain after the polynomial smoothing of the spectra. This unwanted pedestal frequently interferes with the intrinsic shape of the peak and  is very difficult to remove. After testing various techniques, such as flattening or filtering the spectra, we found that the best stability is attained when the peak and pedestal are carefully separated at the cross-correlation function itself. Instead of a blind fitting of a Gaussian function to the peak, we excluded by eye any asymmetrical shoulders from the fit. We also observed that the larger the spectral range, the more symmetrical is the pedestal and the higher the relative amplitude of the peak.
 
\subsubsection{The Stellar Templates}

A set of four template stars was observed with the same instrumental set-up and subsequently used 
to extract both the redshift and velocity dispersion of the galaxies. The
stars are K giants, which contribute the bulk of the light in elliptical galaxies over our observed spectral interval. Their spectral types vary from K0 to K5 and a different degree of matching of the galaxy spectra is expected. To check this, we carried out an internal comparison between the $\sigma$ measurements with the different templates. The reference is the K0 giant SAO 079251 (template B), which showed both a slightly higher than average cross-correlation peak and a smaller velocity error. The results, corresponding to 34 galaxies, are shown in Table 2. The rms scatter of the comparisons with B ( log $\sigma_{\rm templ}$ -- log $\sigma_{\rm templ-B}$ ) for templates A and C is roughly 0.013 in log $\sigma$. Template D, of spectral type K5, showed a larger rms scatter of $\pm$ 0.022 and, as a conservative measure,
 was  excluded from the calculations. Therefore, the final velocity dispersions were averaged from measurements with the A, B and C templates.
 
\placetable{T2}

\subsubsection{The Spectral Range}

We have checked for the dependence of the measured velocity dispersion on the spectral range where the cross-correlation is performed. The following four spectral ranges were tested: (i) $\Delta\lambda_{G}$, around the $G$ band (rest frame 4080--4700 \AA), (ii) $\Delta\lambda_{\rm Mg_{\it b}}$, centered on 
Mg$_{\it }$ (4800--5310 \AA), (iii) $\Delta\lambda_{\rm brack}$, the region bracketed between H$\beta$ and Na D (4900--5800 \AA), and (iv) $\Delta\lambda_{\rm Davies}$, the 4100--6300 \AA\  region
used by Davies et al. (1987;
hereafter D87). The test was carried out using the whole galaxy sample (34 objects) and the three different stellar templates. The internal comparison between the log $\sigma$ values obtained in each spectral range is presented in Table 3, which shows the median offset of log $\sigma_{\rm range 1}$ -- log $\sigma_{\rm range 2} \pm$ the uncertainty on the offset for each stellar template. 

\placetable{T3}

Velocity dispersions measured in both $\Delta\lambda_{G}$ and $\Delta\lambda_{\rm Mg_{\it b}}$ show a significant and systematically large and positive offset, in agreement with the 10 \% value reported by Dressler (1984) for the same two ranges. Differences between $\Delta\lambda_{\rm brack}$ and $\Delta\lambda_{\rm Mg_{\it b}}$ are only marginally significant, perhaps due to the partial match of both spectral ranges. The small offset probably arises from the variability of the H$\beta$ line (due to emission or star formation), which is only present in the $\Delta\lambda_{\rm Mg_{\it b}}$. The significant offset between $\Delta\lambda_{\rm Davies}$ and $\Delta\lambda_{\rm brack}$ is probably due to the different degree of contamination of the peak by the pedestal of the cross-correlation function. As  mentioned above, a large spectral range produces a cross-correlation peak of greater relative amplitude with respect to the pedestal and, consequently, introduces less ambiguity in the peak width measurements and more stable results. The $\Delta\lambda_{\rm Davies}$ is the same spectral range used for the $\sigma$ measurements of our comparison Fundamental Plane (B87). Figure 1 shows one example of the offsets originating from
 the choice of the spectral range, separated according to the reduction method: panels a) and b) correspond to CC and FF methods respectively. The log $\sigma_{\rm Davies}$ -- log $\sigma_{G}$, averaged for the three templates, is represented versus the log $\sigma_{\rm CC}$ value. The lowest-S/N spectra, with S/N $\leq$ 32, were excluded from the representation. The median offset value is significant, with 0.068 $\pm$ 0.008 and an rms scatter of 0.043.

\placefigure{F1}

\subsubsection{Uncertainties}

Prior to the combination of the individual spectra, a trial cross-correlation with the templates was used to correct for both the redshift of the object and the uncertainties in the wavelength scale. In order to assess how the combination process affects the $\sigma$ results, the CC method was applied to all the available spectra, both individual and combined. The differences, log $\sigma_{\rm individual}$ -- log $\sigma_{\rm combined}$, are presented in Figure 2  as a function of the S/N (per \AA) of the individual exposures. The median offset in log $\sigma$ is 0.000 $\pm$ 0.003 with 0.031 rms scatter, indicating that the combination process does not introduce a systematic effect on the derived velocity dispersion. 

The errors for the CC method were estimated from repeated observations of the same objects. Due to the lack of a systematic offset between combined and individual exposures, the error is taken as the rms deviations of the (log $\sigma_{\rm individual}$ -- log $\sigma_{\rm combined}$) around the median value. The rms deviations are calculated in $\Delta$(S/N) = 10 bins and the results are fitted by a function, giving the variation of errors with S/N. Errors generally increase both for lower S/N and lower velocity dispersions, in agreement with the simulations carried out by J\o rgensen et al. (1995b), in which the highest relative systematic errors occur in spectra with both low S/N and low $\sigma$. For the three cases with single observations, formal errors were assigned according to their S/N. The adopted error of an object with S/N = snr$_1$ is the average of the errors in a $\Delta$(S/N) = 10 bin, centered on snr$_1$. 

\placefigure{F2}

\subsection{Fourier-Fitting Method}

\subsubsection{Preparation and Filtering}

Before using the FF method, we  checked for its sensitivity to some of the parameters and procedures in order to reach the best performance. For instance,  continuum subtraction in both the object and template spectra proves to be an important issue. We have checked different options for the continuum fitting, varying the order of the fitting function, the rejection limits, and the spectral range. As a conclusion, the most robust procedure appears to be  local continuum subtraction, {\it i.e.,} only inside the spectral range used for the $\sigma$ calculation, with an 6th-order spline function and a rather low rejection limit (1 in units of the residual sigma). 

After  continuum subtraction, the end points are fixed to zero with a cosine bell, the 
Fourier transforms are calculated and the high and low frequencies
subsequently filtered out. 
We  also checked for the influence of these frequency cut-offs. The high-frequency cut-off ({hfc}) is set at roughly $k_{\rm h} \approx$ (wavelength range)/(spectral resolution). This is a reasonable choice, because our tests show that $\sigma$ results are insensitive to small departures around that hfc value.

The low-frequency components are more difficult to filter. They arise from incomplete continuum subtraction and intrinsic spectral line blends (width 
20--40 \AA), being therefore dependent on the details of each spectral range. The corresponding cut-off is generally chosen to avoid features with frequencies smaller than $\approx$ 100 \AA $^{-1}$ and, therefore, $k_{\rm l} \approx$ (wavelength range)/(100 \AA) [J\o rgensen et al, 1995b]. Other authors (Sargent et al. 1977) propose a $k_{\rm l} = 2/f$, where $f$ is the fraction of the spectrum masked at each end by the cosine bell function. These two prescriptions are not identical. For example, with a 500 \AA\ spectral range and $f = 0.2$, first approach suggests $k_{\rm l}$ = 5, while the second gives $k_{\rm l}$ = 10. 

To test the influence of the choice of the low-frequency cut-off ({lfc}) in the derived $\sigma$, we have applied the FF method to the spectra of five different sample galaxies, using eleven different {lfc} values between 150 and 50 \AA$^{-1}$ and the four different spectral ranges defined in section 3.1.4. Our tests show that the measured $\sigma$ varies with {lfc}. We first normalized the eleven $\sigma$ values, by dividing by their average ($\sigma_{\rm average}$). The shape of the variation of $\sigma$ (lfc) is strongly dependent on the spectral range, as stated in the previous paragraph, and almost independent on the particular galaxy. Therefore, a low order function has been fitted to the set of 55 $\sigma$ values of each spectral range. From Figure 3, which shows the peculiar shape of $\sigma/\sigma_{\rm average}$ in the four spectral ranges, we can conclude that the {lfc} choice can both minimize or exacerbate the internal inconsistencies between spectral ranges. For instance, {lfc} = (70-75) \AA $^{-1}$ would make the best bet, while {lfc} = 100 or 60 \AA $^{-1}$, where the variations are ``out of phase'', would worsen the internal inconsistencies between spectral ranges. From Figure 3 one can easily deduce that a choice of {lfc} $\approx$ 100 \AA $^{-1}$ would add an offset of log $\sigma_{\rm Davies}$ -- log $\sigma_{G} \approx$ 0.048, against $\approx$ -- 0.022, for {lfc} $\approx$ 75 \AA $^{-1}$. It must be emphasized that, although these results are only valid for our particular continuum subtraction, a search for the best {lfc} value is a worthwhile procedure to decrease the systematic errors in $\sigma$. 

\placefigure{F3}

\subsubsection{Calculations}

The choice of the best {lfc} value, discussed in the previous section, does not guarantee a perfect internal consistency in the $\sigma$ results obtained with different spectral ranges. In despite of the use of an optimum {lfc} = 75 ($k_{\rm l}$ = 29.3) there is a remaining median offset log $\sigma_{\rm Davies}$ -- log $\sigma_{G}$ = 0.047 $\pm$ 0.013, with an rms scatter of 0.071, when calculations are carried out with the FF method (see Figure 1b). Our results are in contradiction with the finding of J\o rgensen et al. (1999), whose median offset in log $\sigma_{\rm Mg_{\it b}}$ -- log $\sigma_{G}$ = 0.00 $\pm$ 0.02. For these same ranges we find a median offset = 0.051  $\pm$ 0.016, with an rms scatter of 0.088. Thus, the offset is significant on the 3$\sigma$ level. One possible source of the disagreement is our poor S/N for $\Delta\lambda_{G}$, noticeably smaller than in $\Delta\lambda_{\rm Davies}$. Similarly to the CC case (section 3.1.4), our $\sigma_{\rm FF}$ values are also calculated over the spectral range $\Delta\lambda_{\rm Davies}$.

\subsubsection{Uncertainties}
Figure 4 represents the formal relative uncertainties of the FF method versus the S/N per \AA . In order to keep the uncertainty below 10 \%, spectra with S/N $<$ 45 are excluded from the present study and marked with an asterisk in Table 5. Galaxies with log $\sigma \leq 2.0$ were automatically excluded, because they had poor S/N, except for NGC 0221 (M32) whose log $\sigma$ (= 1.97) is accompanied by a large S/N (= 247) and a low uncertainty. 
  
\placefigure{F4}

When both methods are compared, the remaining sample (24 galaxies), with S/N $>$ 45, shows a nonsignificant median offset in $\Delta$ log $\sigma_{\rm CC -- FF}$ = log $\sigma_{\rm CC}$ -- log $\sigma_{\rm FF}$ = --0.004 $\pm$ 0.006, with an rms scatter of 0.029. This rms scatter around the median value is of the order of the quadratic sum of the internal uncertainties for both methods. Uncertainties for the CC method are typically 0.014 dex, measured with repeated observations, and 0.020 dex for FF, deduced from the formal uncertainties given by the Fourier Fitting method.

Figure 5 shows the result of the internal comparison between both CC and FF methods, for the S/N $>$ 45 sample, with log $\sigma_{\rm CC}$ -- log $\sigma_{\rm FF}$ versus log $\sigma_{\rm FF}$. Error bars represent the CC and FF uncertainties added in quadrature. The lack of a systematic offset in $\Delta$ log $\sigma_{\rm CC -- FF}$ allows us to choose between any of the two methods to calculate $\sigma$. We have preferred the FF method because it offers a more reliable error analysis than CC. From now on, our $\sigma$ results are, consequently, $\sigma_{\rm FF}$.

\placefigure{F5}

\section{External Comparison}

\subsection{The Field Sample}

The goal of the present study is to compare the FP for galaxies in the Hickson compact groups with the FP for galaxies in other environments. It is, therefore, of paramount importance to establish how different templates and reduction methods affect the derived $\sigma$. With this purpose in mind, we compared our final velocity-dispersion values ($\sigma_{\rm FF}$) with data published in the literature. Fortunately, the twelve galaxies (NGCs with S/N $ > $ 45) of our field control sample, have been repeatedly observed by other authors. We have selected two different sources for the external comparison: the compilation by McElroy (1995) and the observations by D87. The first source updates the former Whitmore, McElroy, \& Tonry (1985) catalog of published velocity dispersions. Their compiled $ \sigma$ values are calculated through weighted averages of several observations, which had been previously standardized with scaling factors and reduced to a common aperture ($2^{\prime\prime} \times 4^{\prime\prime} $). The main advantage of this compilation is that data averaged from multiple normalized sources are less prone to  systematic errors due to a particular telescope/detector configuration. Futhermore, 7 out of 12 of our sample galaxies are included in the so called {\it standard galaxies} subset, defined by McElroy (1995) as those having at least three reliable, concordant measurements.

We have extracted a homogeneous subset from the second comparison source, D87, by selecting their Lick 3-m telescope/IDS observations. Despite having a poor instrumental dispersion (215 km s$^{-1}$), there are two main advantages in the use of this comparison source. On the one hand, both our sample and D87--Lick subset use the same spectral range and, on the other, all our sample galaxies are included in the D87--Lick subset. We have worked out average $\sigma$ values from the published D87--Lick measurements. The D87 data set is one among several contributions to the compilation by McElroy (1995) and, therefore, an absolute independence of our two external sources cannot be guaranteed.  

In order to proceed with the comparison, all the raw velocity dispersions have to be aperture-corrected to the same metric radius. Following the study by J\o rgensen et al. (1995b), the aperture correction is applied via the power law that best fits the variation of $\sigma$ with aperture for a set of empirical models. The power law  $${\rm log}\left(\sigma_{\rm ap}\over \sigma_{\rm norm}\right) = -- 0.04\ {\rm log}\left(r_{\rm ap}\over r_{\rm norm}\right)$$ shows a very good compromise, with the best performance in the interval: log (r$_{\rm ap}$/r$_{\rm norm}) \approx$ --0.6 to + 0.6, which roughly includes 67 \% of our NGC data, 50 \% from McElroy (1985) and 42 \% from D87.

Our log($r_{\rm ap}/r_{\rm norm}$) data are listed in column 10 of Table 1. Data from McElroy (1995), normalized to a rectangular aperture of $2^{\prime\prime}\times 4^{\prime\prime}$,  have an equivalent circular radius $r_{\rm ap}$ = 1.64$^{\prime\prime}$.  Lick data from D87 with $1.5^{\prime\prime}\times 4^{\prime\prime}$ have $r_{\rm ap}$ = 1.42 arcsec. Their corresponding log ($r_{\rm ap}/r_{\rm norm}$) ratios are calculated with the distances of the galaxies and $r_{\rm norm}$ = 0.595 $h^{-1}$ kpc.

The median value for the aperture corrections, log ($\sigma_{\rm ap}$/ $\sigma_{\rm norm}$), is 0.022 for our observations, 0.027 for McElroy (1995) and 0.030 for D87.

The external comparison of our data with the literature is presented in Figure 6 (panels a and b) and Table 4, which shows the median averages of log $\sigma_{\rm ours}$ -- log $\sigma_{\rm literature}$. The offsets are not significant and the residual scatters roughly agree with expectations from reported errors. However, there is a worrying systematic trend in the comparison with D87 (Figure 6b), with larger offsets (in the sense of $\sigma_{\rm Lick}$ being smaller) for lower $\sigma$ values.  There are reasons to suspect that the origin of this effect is not in our data, but in the poor instrumental dispersion of the D87--Lick subsample, because a similar systematic deviation appears when the same D87--Lick subsample is compared with the other literature source (McElroy 1995; Figure 6 panel c). Nevertheless, other authors using a larger sample (J\o rgensen et al. 1995b) and D87, in their internal comparison, do not find such a trend with respect to the Lick subsamples. We shall take our trend into account during the comparison with the B87 Fundamental Plane, which uses the D87 velocity dispersions.

\placetable{T4}

\placefigure{F6}

\subsection{The HCG Sample}

An extra source of external comparison is the spectroscopic study of elliptical galaxies in Hickson compact groups (HCGs) by ZW93, which has
 ten $\sigma$ measurements in common with the present study (only those with S/N $>$ 45). The aperture correction used by ZW93 is similar to that of D87 and, therefore, our comparison sample must also follow the same correction procedure. To this purpose, our data have been extracted from the   log $\sigma_{\rm ap/D87}$ column in Table 5. The result of the comparison is presented in Figure 6 (panel d) and Table 4. After the exclusion of galaxy HCG 44b, which ZW93 took from the B87 sample, there is a significant median offset of 0.059 $\pm$ 0.029. We interpret this offset as almost certainly produced by the use of different spectral ranges for $\sigma$ derivations. In the ZW93 subsample, KPNO and CTIO$_1$ spectra, corresponding to 90 \% of the observations, have spectral ranges (4100--4542 and 4060--4575 \AA) 
shorter even than the $\Delta\lambda_{G}$ discussed in Section
 3.1.4. It was shown that systematic offsets arise from the use of different ranges. For instance, the difference log $\sigma_{\rm Davies}$ -- log $\sigma_{G}$ (averaging for the three stellar templates in Table 3) amounts to 0.054, almost identical to the log $\sigma_{\rm ours}$ -- log $\sigma_{\rm ZW93}$ in Table 4.

\section{Comparison of the Fundamental Plane for Different Samples}

\subsection{Corrections and Results}
\subsubsection{Corrections}

The goal of the present study is the comparison of the FP for our sample of galaxies in HCGs with the FP found by B87 for a large sample of galaxies in different environments, whose source of velocity dispersion data is D87. In an attempt to reproduce their $\sigma$ determination, we have used their same aperture correction (formula 1 in D87), with $V_{\rm Virgo}$ = 1100 km s$^{-1}$ and $V_{\rm Coma}$ = 7000 km s$^{-1}$. The D87 aperture correction is slightly different from that proposed in J\o rgensen et al (1995b;  see differences in their Figure 4c) used by us in Section 4.1. A further correction is related to the trend discovered in our external comparison with the D87--Lick subsample. We have re-calculated the trend with the proper D87 sample, whose {\it adopted} $\sigma$ is the average of several observational sources. A least-squares fit of the trend gives:  
log $\sigma_{\rm ours}$ -- log $\sigma_{\rm D87}$ = 0.43 -- 0.19 $\times$ log $\sigma_{\rm ours}$, where $\sigma_{\rm ours}$ is the $\sigma_{\rm FF}$ with the D87 aperture correction. The large contribution of NGC 221 (M32) to the measured trend (e.g. Figure 6b) is probably due to the combination of the small $\sigma_{\rm ours} \approx$ 79 km s$^{-1}$ of the galaxy with the poor Lick/IDS instrumental resolution (215 km s$^{-1}$). Nevertheless, galaxy NGC 221 has been kept for the trend calculations, because its removal from the sample does not significantly change the slope. 

\subsubsection{Results}

Table 5 contains the $\sigma$ results of the present study in addition to other two parameters used to define the Fundamental Plane. Galaxies marked with an asterisk, in column (1), have an S/N $<$ 45 and were accordingly excluded from the FP. Column (2) shows raw log $\sigma_{\rm FF}$ and column (3) its uncertainty. Column (4) presents the previous log $\sigma_{\rm FF}$ corrected from aperture in the {\it J95-fashion} (log $\sigma_{\rm ap/J95}$). Column (5) has the log $\sigma_{\rm FF}$ corrected from aperture in the {\it D87-fashion} (log $\sigma_{\rm ap/D87}$), to be used in the comparison with the B87 Fundamental Plane. Column (6) shows the log $\sigma_{\rm ap/D87}$ values corrected for the trend detected in the comparison with D87, according to the relation log $\sigma_{\rm ap+tr}$ = 1.19 $\times$ log $\sigma_{\rm ap/D87}$ -- 0.43 (see previous section). The other characteristic parameters, $\langle \mu_{B} \rangle_{\rm eff}$ (column 7) and log $r_{\rm eff}$ (column 8), were gathered from the literature. Data for the HCG sample come from ZW93, except for HCG 93a which comes from Zepf (1991). Those for the NGC galaxies were extracted from B87.

\placetable{T5} 

Out of the 34 originally observed galaxies included in Table 1, the
four with the poorest S/N were excluded from the list of results (Table 5) because reduction methods failed to provide a reliable $\sigma$ result. Out of the remaining 30 objects, four lacked photometric data in the literature and three more objects with S/N $<$ 45 were also excluded from the FP. In summary, we have ended up with a sample of 11 HCG and 12 NGC galaxies to carry out the FP comparison.

\subsection{The Fundamental Plane}

\subsubsection{Comparison with the Control FP}

The standard FP equation is usually written as log $r_{\rm eff}$ = a log $\sigma$ + b $\langle \mu_{B} \rangle_{\rm eff}$ + c, where $a$ and $c$ are respectively referred to as the slope and the intercept of the FP. We have adopted the $b$ = 0.32 value which the majority of the studies of the FP have obtained for nearby clusters, independently of the fitting method or wavelength range. For instance, $b_{R_C}$ = 0.326 $\pm$ 0.011 (Hudson et al. 1997); $b_{Gunn r}$ = 0.328 $\pm$ 0.008 (J\o rgensen et al. 1996); $b_{I_C}$ = 0.320 $\pm$ 0.012 (Scodeggio et al. 1997) and $b_K$ = 0.315 $\pm$ 0.011 (Pahre et al. 1998a).

The control FP is constructed from the large, homogeneous sample of elliptical galaxies in B87. Out of the original B87 sample of 456 galaxies, we excluded galaxies with $\sigma \leq$ 100 km s$^{-1}$, $cz <$ 1000 km s$^{-1}$, $M_{B} \geq$ --18, and those without  $(B - V)_0$ color. The FP of the remaining sample of 339 galaxies follows the log $r_{\rm eff}$ = 0.98 log $\sigma$ + 0.32 $\langle \mu_{B} \rangle_{\rm eff}$ - 5.57 relation. Figure 7 represents an edge-on perspective of the FP with log $r_{\rm eff}$ versus log $\sigma$ + 0.326 $\langle \mu_{B} \rangle_{\rm eff}$, where the dashed line represents the fitted control FP. 

The largest points in Figure 7 correspond to our sample galaxies with S/N $>$ 45, discriminated according to their environment: NGC field galaxies are represented by solid triangles and HCG galaxies by solid circles.  The log $\sigma$ values are corrected for both the aperture in the {\it D87-fashion} and for the trend detected in the external comparison (column log $\sigma_{\rm ap+tr}$ in Table 5). Note that galaxy NGC 4552 (M 89) shows a large departure from the FP. This departure is probably due to problems with the surface photometry of the galaxy. In a study by Caon et al (1990) it is shown that this Virgo galaxy has an odd luminosity profile with a change of slope which is typical of tidally distorted ellipticals. For instance, the $r_{\rm eff}$ measured by Caon et al (1990) is more than three times larger than the one measured by B87.

The numeric results of the comparison with the control FP are presented in Table 6, using the format $Y_{\rm ours} - Y_{\rm control}$, where $Y$ = log $\sigma$ + 0.326 $\langle \mu_{B} \rangle_{\rm eff}$. The ZW93 values are calculated in a similar way, using their published values for: $\sigma$, $\langle \mu_{B} \rangle_{\rm e}$ and $r_{\rm e}$.

\placefigure{F7}

\subsubsection{Uncertainties}

Errors for the spectroscopic parameter of the FP ($\sigma$) are found in column $\sigma _{\rm log \sigma}$ of Table 5, while those for the photometric parameters are discussed in the corresponding literature sources (ZW93 and B87). However, the strong coupling between $\langle \mu_{B} \rangle_{\rm eff}$ and log $r_{\rm eff}$ conspires in such a way that correlated errors tend to produce movements parallel to the FP rather than offsets from it (see  discussion in ZW93). In Figure 7, we have therefore ignored the photometric uncertainties and only included error bars corresponding to errors in log $\sigma$.

\placetable{T6}

\subsubsection{The FP Face-on}

In the face-on view of the FP (Figure 8) both the HCG galaxies and the B87 comparison sample populate the same region of the plane, reinforcing the conclusion that both samples are essentially similar. In this projection, $\langle I \rangle_{\rm e}$ is related to $\langle \mu_{B} \rangle_{\rm eff}$ by $\langle \mu_{B} \rangle_{\rm eff}$ = -- 2.5 log $\langle I \rangle_{\rm e}$ and r$_{\rm eff}$ is measured in kiloparsecs. As pointed out by Guzm\'an, Lucey \& Bower (1993), the galaxies only populate a small elongated area of the available plane. The lower-left boundary is probably a selection effect resulting from the criteria imposed on the B87 sample, i.e. $\sigma >$ 100 km s$^{-1}$ and $M_{B} <$ --18. The upper-right boundary, on the contrary, is a physical effect which could be caused by the lack of galaxies with $\sigma >$ 300 km s$^{-1}$ or by the lack of very luminous galaxies with high $\langle \mu_{B} \rangle_{\rm eff}$.  

As a reference, we have plotted in Figure 8 the lines which enclose our sample HCG galaxies and the majority of the comparison sample. Points are found inside the region limited by 20.3 $\leq \langle \mu_{B} \rangle_{\rm eff} \leq$ 23.0 mag arcsec$^{-2}$ (dashed lines) and 120 $\leq \sigma \leq $ 300 km s$^{-1}$ (dotted lines). The properties of our HCG sample galaxies vary from HCG 37a (coordinates: 5.6, -5.5), the most diffuse object, to HCG 44b (4.5, -5.0) and HCG 10b (4.7, -4.8), the most compact ones, with a relatively small $r_{\rm eff}$ and the brightest $\langle \mu_{B} \rangle_{\rm eff}$).

\placefigure{F8}
  
\section{Discussion}

After the application of both the aperture and trend corrections, we may conclude from Table 6 that, from the point of view of the Fundamental Plane, there is little or no significant 
difference between elliptical galaxies in HGCs and comparable galaxies in other environments. This does not support the previous finding by ZW93, who reported $\sim$ 20 \% lower velocity dispersions in E galaxies of CGs than in their counterparts in other environments. In \S  4.2 we interpreted this disagreement in terms of the spectral range used to determine $\sigma$ in ZW93, quite different from that used in the control sample (D87) and in the present paper. This interpretation is reinforced on checking the behavior of the three CG galaxies (HCG 22a, HCG 42a, and HCG 44b), 
whose data ZW93 extracted directly from the B87 sample. Although they only represent 14 \% of the whole ZW93 sample, they should also show traces of the $\sigma$ deficit reported in their counterparts. On the contrary, the median offset of those three galaxies with respect to the FP, $Y_3 - Y_{\rm control}$ = 0.011  $\pm$ 0.044 is insignificant, in contrast to the --0.074  $\pm$  0.024 of the whole ZW93 sample. The present study has stressed the reproducing, as far as possible, of the reduction and corrections used in the control sample to avoid those unwanted systematic effects. 
   
The main result of this present contribution is that, by gathering higher-quality data and getting rid of systematic errors, we were able to measure more reliably the FP of ellipticals in CGs, and only a marginally significant difference was observed when compared to the FP of ellipticals in various environments. This information adds to the recent result by Pahre,
Djorgovski, \& de Carvalho (1998a,b),
 who found no difference between ellipticals in clusters and the
field, as far as the FP is concerned, indicating that their dynamical properties are not influenced
by the environment, unless dynamical differences are offset by 
stellar-population differences that act in the opposite direction. We shall address this problem in a study (now in progress) of the stellar populations of the same sample galaxies. In it, the same observational material is used to determine the spectral indices and their relation with other parameters, 
such as the environment or the velocity dispersion. 

\acknowledgments

SEZ acknowledges support from the Hellman Family Fellowship.

\begin{deluxetable}{lccccccccc}
\footnotesize
\tablecaption{Log of the observations \label{T1}}
\tablewidth{0pt}
\tablehead{
Name & Type & 01/09 & 01/10 & 01/11 & 01/12 & $t_{\rm exp}$ & $m_{B}$ & S/N per \AA\ & log ($r_{\rm ap}/r_{\rm norm}$) \nl
}
\startdata							   	
HCG 10b & E1 & 2 $\times$ 600    &        &        & 2 $\times$ 2400 & 6000 & 12.70 & 136 & --0.046 \nl   
HCG 14b & E5 & 2 $\times$ 900    &        &        &        & 1800 & 14.17 & 39  &  0.009 \nl   
HCG 15b & E0 &          & 2 $\times$ 900  &        &        & 1800 & 14.74 & 47  &  0.105 \nl   
HCG 15c & E0 &          & 2 $\times$ 900  &        &        & 1800 & 14.37 & 48  &  0.105 \nl   
HCG 19a & E2 & 2 $\times$ 900    &        & 2 $\times$ 2400 & 2 $\times$ 2400 & 11400& 14.00 & 162 & --0.103 \nl  
HCG 28b & E5 &          & 2 $\times$ 900  &        &        & 1800 & 15.31 & 28  &  0.327 \nl   
HCG 32a & E2 & 2 $\times$ 900    &        &        &        & 1800 & 13.80 & 38  &  0.356 \nl   
HCG 37a & E7 & 3x600    &        & 2 $\times$ 2700 & 2 $\times$ 1350 & 9900 & 12.97 & 131 &  0.095 \nl  
HCG 37e & E0 &          & 2 $\times$ 1200 &        &        & 2400 & 16.21 & 23  &  0.095 \nl   
HCG 40a & E3 & 2 $\times$ 900    &        & 2 $\times$ 2700 & 2 $\times$ 1350 & 9900 & 13.44 & 142 &  0.095 \nl   
HCG 44b & E2 & 2 $\times$ 600    &        &        &        & 1200 & 11.62 & 48  & --0.590 \nl  
HCG 46a & E3 & 2 $\times$ 1200   & 2 $\times$ 900  &        &        & 4200 & 16.40 & 32  &  0.178 \nl  
HCG 46c & E1 &          & 2 $\times$ 1200 &        &        & 2400 & 16.13 & 37  &  0.178 \nl  
HCG 51a & E1 &          &        & 2 $\times$ 900  &        & 1800 & 13.89 & 41  &  0.158 \nl   
HCG 57c & E3 &          & 2 $\times$ 900  &        & 2 $\times$ 2700 & 7200 & 14.63 & 73  &  0.230 \nl   
HCG 57f & E4 &          & 2 $\times$ 900  &        & 2 $\times$ 2700 & 7200 & 15.22 & 70  &  0.230 \nl   
HCG 59b & E0 &          & 2 $\times$ 1200 &        &        & 2400 & 15.20 & 32  & --0.123 \nl  
HCG 62a & E3 &          & 2 $\times$ 600  &        &        & 1200 & 13.36 & 47  & --0.116 \nl  
HCG 68b & E2 &          & 2 $\times$ 300  &        &        & 600  & 12.24 & 39  & --0.350 \nl
HCG 93a & E1 & 2 $\times$ 900    &        &        &        & 1800 & 12.61 & 60  & --0.028 \nl  
HCG 96b & E2 &          & 2 $\times$ 900  &        &        & 1800 & 14.10 & 50  &  0.212 \nl   
HCG 97a & E5 &          &        &        & 2 $\times$ 600  & 1200 & 14.16 & 35  &  0.085 \nl
\tableline   
NGC 221 & E2 & 2 $\times$ 360    & 2 $\times$ 120  &        &        & 960  & 9.03  & 247 & --1.827 \nl  
NGC 584 & E4 &          & 2 $\times$ 300  &        & 2 $\times$ 300  & 1200 & 11.44 & 100 & --0.458 \nl  
NGC 636 & E3 & 2 $\times$ 450    &        & 2 $\times$ 450  &        & 1800 & 12.41 & 89  & --0.463 \nl  
NGC 821 & E6 &          & 2 $\times$ 450  &        &        & 900  & 11.67 & 70  & --0.495 \nl  
NGC 1700 & E4 & 2 $\times$ 450    & 2 $\times$ 450  & 2 $\times$ 300  & 2 $\times$ 300  & 3000 & 12.20 & 129 & --0.141 \nl  
NGC 2300 & SA0 & 2 $\times$ 450   & 2 $\times$ 450  & 2 $\times$ 300  & 2 $\times$ 300  & 3000 & 12.07 & 89  & --0.447 \nl  
NGC 3377 & E5 & 2 $\times$ 450    & 2 $\times$ 300  & 2 $\times$ 300  &        & 2100 & 11.24 & 120 & --0.974 \nl  
NGC 3379 & E1 & 2 $\times$ 300    &        &        &        & 600  & 10.24 & 62  & --0.766 \nl  
NGC 4552 & E0 &          & 2 $\times$ 300  &        &        & 600  & 10.73 & 78  & --1.225 \nl
NGC 4649 & E2 &          & 1 $\times$ 300  &        &        & 300  &  9.81 & 48  & --0.580 \nl
NGC 4697 & E6 &          & 2 $\times$ 300  & 2 $\times$ 300  &        & 1200 & 10.14 & 92  & --0.637 \nl  
NGC 7619 & E2 &          & 2 $\times$ 450  &        &        & 900  & 12.10 & 62  & --0.150 \nl
\enddata  
\end{deluxetable}

\begin{deluxetable}{lcccc}
\footnotesize
\tablecaption{Template comparison \label{T2}}
\tablewidth{0pt}
\tablehead{
Template & Spec. type & $\langle\Delta \log \sigma\rangle$ & rms$\langle\Delta \log \sigma\rangle$ & $\langle$height$_{\rm peak} \rangle$ \nl
 & & log $\sigma_{\rm templ}$ -- log $\sigma_{\rm templ-B}$ & & \nl
}
\startdata
A: HD  030104 & K0 &  0.001 & 0.014 & 0.847 \nl
B: SAO 079251 & K0 & .....  & ..... & 0.851 \nl
C: SAO 150504 & K2 & --0.011& 0.012 & 0.827 \nl
D: SAO 150608 & K5 &  0.019 & 0.022 & 0.829 \nl
\enddata 
\end{deluxetable}

\begin{deluxetable}{lcccccc}
\footnotesize
\tablecaption{Spectral-range comparison \label{T3}}
\tablewidth{0pt}
\tablehead{
Range & Template A & & Template B & & Template C & \nl
 & median & rms scatter & median & rms scatter & median & rms scatter \nl}
\startdata
log $\sigma_{\rm Mg_{\it b}}$ -- log $\sigma_{G}$       &  0.176 $\pm$ 0.015 & 0.09 & 0.110 $\pm$ 0.014 & 0.08 & 0.140 $\pm$ 0.015 & 0.09  \nl
log $\sigma_{\rm brack}$ -- log $\sigma_{\rm Mg_{\it b}}$   & --0.038 $\pm$ 0.010 & 0.06 & --0.022 $\pm$ 0.010 & 0.06 & --0.030 $\pm$ 0.014 & 0.08  \nl
log $\sigma_{\rm Davies}$ -- log $\sigma_{\rm brack}$& --0.060 $\pm$ 0.010 & 0.06 & --0.050 $\pm$ 0.009 & 0.05 & --0.063 $\pm$ 0.009 & 0.05   \nl
\enddata 
\end{deluxetable}

\begin{deluxetable}{ccc}
\footnotesize
\tablecaption{Comparison (ours -- literature) \label{T4}}
\tablewidth{0pt}
\tablehead{
$\langle\Delta \log \sigma\rangle$ & $\langle\Delta \log \sigma\rangle$ & $\langle\Delta \log \sigma\rangle$ \nl
D87 & McElroy (1995) & ZW93 \nl
}
\startdata
0.002  $\pm$  0.011 & 0.005  $\pm$  0.010 & 0.059  $\pm$  0.029 \nl
\enddata 
\end{deluxetable}

\begin{deluxetable}{lccccccc}
\footnotesize
\tablecaption{$\sigma_{\rm FF}$ results and FP parameters \label{T5}}
\tablewidth{0pt}
\tablehead{
Galaxy & log $\sigma_{\rm FF}$ & $\sigma _{\rm log \sigma}$ & log $\sigma_{\rm ap/J95}$ & log $\sigma_{\rm ap/D87}$ &  log $\sigma_{\rm ap+tr}$ & $\langle \mu_{B} \rangle_{\rm eff}$ & log $r_{\rm eff}$ (pc) \nl
}
\startdata
HCG 10b      & 2.390 & 0.011 & 2.388 & 2.382 & 2.404 &  20.54 & 3.42 \nl  
HCG 14b$\ast$& 1.983 & 0.074 & 1.983 & 1.977 & 1.923 &  24.06 & 4.08 \nl 
HCG 15b      & 2.135 & 0.040 & 2.139 & 2.135 & 2.110 &  21.18 & 3.44 \nl 
HCG 15c      & 2.213 & 0.027 & 2.209 & 2.212 & 2.202 &  21.42 & 3.53 \nl  
HCG 19a      & 2.249 & 0.008 & 2.262 & 2.239 & 2.234 &  20.84 & 3.27 \nl 
HCG 32a$\ast$& 2.274 & 0.060 & 2.278 & 2.293 & 2.299 &  22.36 & 3.96 \nl  
HCG 37a      & 2.416 & 0.012 & 2.420 & 2.415 & 2.444 &  22.76 & 4.12 \nl 
HCG 40a      & 2.405 & 0.010 & 2.381 & 2.404 & 2.430 &  21.38 & 3.68 \nl  
HCG 44b      & 2.256 & 0.031 & 2.263 & 2.235 & 2.230 &  20.69 & 3.20 \nl 
HCG 46c$\ast$& 1.973 & 0.120 & 1.980 & 1.977 & 1.923 &        &      \nl 
HCG 51a$\ast$& 2.241 & 0.075 & 2.250 & 2.244 & 2.240 &        &      \nl  
HCG 57c      & 2.330 & 0.019 & 2.339 & 2.338 & 2.352 &  22.55 & 3.85 \nl  
HCG 57f      & 2.150 & 0.023 & 2.152 & 2.158 & 2.138 &  22.27 & 3.67 \nl  
HCG 62a      & 2.340 & 0.035 & 2.326 & 2.329 & 2.341 &        &      \nl 
HCG 68b$\ast$& 2.219 & 0.052 & 2.205 & 2.202 & 2.190 &        &      \nl 
HCG 93a      & 2.358 & 0.023 & 2.372 & 2.351 & 2.368 &  21.56 & 3.71 \nl
HCG 96b      & 2.289 & 0.027 & 2.298 & 2.296 & 2.302 &  20.71 & 3.44 \nl 
HCG 97a$\ast$& 2.131 & 0.065 & 2.134 & 2.129 & 2.104 &  23.27 & 4.13 \nl 
\tableline
NGC 221       & 1.977 & 0.015 & 1.904 & 1.950 & 1.891 & 18.69  & 2.18 \nl
NGC 584       & 2.313 & 0.010 & 2.295 & 2.294 & 2.300 & 20.44  & 3.40 \nl
NGC 636       & 2.228 & 0.011 & 2.209 & 2.208 & 2.198 & 20.71  & 3.23 \nl
NGC 821       & 2.302 & 0.016 & 2.282 & 2.282 & 2.286 & 21.85  & 3.58 \nl
NGC 1700      & 2.383 & 0.010 & 2.378 & 2.373 & 2.393 & 19.94  & 3.41 \nl
NGC 2300      & 2.415 & 0.017 & 2.397 & 2.396 & 2.421 & 21.51  & 3.53 \nl
NGC 3377      & 2.171 & 0.014 & 2.132 & 2.147 & 2.125 & 20.76  & 2.97 \nl
NGC 3379      & 2.335 & 0.019 & 2.304 & 2.311 & 2.320 & 20.16  & 3.20 \nl
NGC 4552      & 2.435 & 0.021 & 2.386 & 2.413 & 2.442 & 20.22  & 2.67 \nl
NGC 4649      & 2.517 & 0.033 & 2.493 & 2.495 & 2.539 & 21.1   & 3.70 \nl
NGC 4697      & 2.236 & 0.015 & 2.211 & 2.215 & 2.205 & 21.41  & 3.66 \nl
NGC 7619      & 2.468 & 0.022 & 2.462 & 2.455 & 2.491 & 21.53  & 3.77 \nl
\enddata 
\end{deluxetable}

\begin{deluxetable}{lcccc}
\footnotesize
\tablecaption{Comparison with the Fundamental Plane \label{T6}}
\tablewidth{0pt}
\tablehead{ &  $Y_{\rm NGC}$ -- Y$_{\rm B87}$ & $Y_{\rm HCG}$ -- Y$_{\rm B87}$ \nl
}
\startdata
Aperture correction only & --0.034  $\pm$  0.069 & --0.060  $\pm$  0.022 \nl
Aperture and trend corrections &  --0.022  $\pm$  0.069 & --0.033  $\pm$  0.023 \nl
ZW93 (their full sample) & ... & --0.074  $\pm$  0.024 \nl
ZW93 (same as our subsample) & ... & --0.103 $\pm$  0.028 \nl
\enddata 
\end{deluxetable}

\clearpage

\figcaption[fig1.ps]{
Internal comparison between $\sigma$ derived from different spectral ranges. As an example, log $\sigma_{\rm Davies}$ -- log $\sigma_{G}$ is represented versus log $\sigma_{CC}$. Panels a) and b) show $\sigma$ results obtained through respectively CC and FF methods. Error bars correspond to the internal uncertainties of each method. \label{F1}}

\figcaption[fig2.ps]{
The comparison between $\sigma$ derived (via the CC method) from individual frames and combined spectra is used to assess possible systematic effects. A null offset excludes that possibility. The rms deviation of the points is used to deduce errors in the $\sigma_{\rm CC}$. \label{F2}}

\figcaption[fig3.ps]{
Variation of the derived normalized $\sigma$ (via the FF method) with the 
low-frequency cut-off (lfc) of the spectra, used during filtering. Spectral details produce a different variations for each spectral range. Each line corresponds to the best fit to the results of five different spectra. Internal consistency in the derived $\sigma$ improves when an lfc $\approx$ 75 is used. \label{F3}}

\figcaption[fig4.ps]{
Formal relative uncertainties, given by the FF method, versus the S/N per \AA. Spectra with S/N per \AA\ $>$ 45 are needed to keep the uncertainty below the 10 \% level. \label{F4}}

\figcaption[fig5.ps]{
Internal comparison of the $\sigma$ obtained with the CC and FF methods. 
Only the subsample with S/N $>$ 45 is used in the presentation. 
Error bars represent the CC and FF uncertainties added in quadrature. \label{F5}}

\figcaption[fig6.ps]{
External comparison with the literature. The format is log $\sigma_{\rm ours}$ 
-- log $\sigma_{\rm literature}$, (except for panel c) versus the final $\sigma_{\rm FF}$. Panels a) and b) show the comparison with McElroy (1995) and the D87--Lick subsample. Panel c) compares two literature sources (McElroy 1995--D87) in support of our opinion that the origin of the trend in panel b) is in the  D87 data. Panel d) shows comparison of our HCG sample with ZW93. \label{F6}} 

\figcaption[fig7.ps]{
Plot of the Fundamental Plane (FP) of elliptical galaxies. The small points and  their fitted dashed line represent the control FP from B87 and 
the large points are the results of the present study. The symbols discriminate different environments: {\it triangles} are field galaxies (NGC subsample) and {\it circles} represent elliptical galaxies in Hickson compact groups (HCG subsample). Error bars only represent uncertainties in log $\sigma$, because errors in the photometric parameters are correlated and do not contribute to the offset. A qualitative diagram in the lower right-hand corner shows how photometric errors move objects parallel to the FP. \label{F7}}

\figcaption[fig8.ps]{
The Fundamental Plane (FP) seen face-on, displaying our sample HCG galaxies and the comparison B87 sample. Lines drawn in the upper-right corner show the directions in which the observed quantities vary along the plane. All our sample HCG galaxies and the majority of those in the comparison sample are found to have 20.3 $\leq \langle \mu_{B} \rangle_{\rm eff} \leq$ 23 mag arcsec$^{-2}$ (dashed lines) and 120 $\leq \sigma \leq $ 300 km s$^{-1}$ (dotted lines). Both samples only populate a small elongated area of the plane. The lower-left boundary is probably a selection effect, while the upper-right boundary is a physical effect caused by the existence of some constraints on the galaxy structure. \label{F8}}   


\begin{references}

\reference{} Burstein, D., Davies, R. L., Dressler, A., Faber, S. M., Stone, R. P. S., Lynden-Bell, D., Terlevich, R. J., \& Wegner, G. 1987, \apjs, 64, 601 (B87)
\reference{} Caon, N., Capaccioli, M., \& Rampazzo, R. 1990, A\&AS, 86, 429
\reference{} Dalle Ore, C., Faber, S. M., Jes\'us, J., \& Stoughton, R. 1991, \apj, 366, 38
\reference{} Davies, R. L., Burstein, D., Dressler, A., Faber, S. M., Lynden-Bell, D., Trelevich, R. J., \& Wegner, G. 1987, \apjs, 64, 581 (D87)
\reference{} de Carvalho, R. R., \& Djorgovski, S. 1992, \apj, 389, L49
\reference{} de Vaucouleurs, G., de Vaucouleurs, A., Corwin, Jr, H. G., Buta, R. J., Paturel, G., \& Fouqu\'e, P. 1991, Third Reference Catalogue of Bright Galaxies (New York: Springer)
\reference{} Djorgovski, S., \& Davis, M. 1987, \apj, 313, 59
\reference{} Dressler, A. 1984, \apj, 281, 512
\reference{} Dressler, A., Lynden-Bell, D., Burstein, D., Davies, R. L., Faber, S. M., Terlevich, R. J., \& Wegner, G. 1987, \apj, 313, 42
\reference{} Faber, S. M., Wegner, G., Burstein, D., Davies, R. L., Dressler, A., Lynden-Bell, D., \& Terlevich, R. J. 1989, \apjs, 69, 763
\reference{} Franx, M., Illingworth, G., \& Heckman, T. 1989, \apj, 344, 613
\reference{} Guzm\'an, R., Lucey, J. R., \& Bower, R. G. 1993, \mnras, 265, 731
\reference{} Guzm\'an, R., Lucey, J. R., Carter, D., \& Terlevich, R. J. 1992, \mnras, 257, 187
\reference{} Hickson, P. 1982, \apj, 255, 382
\reference{} Hickson, P. 1994, Atlas of Compact Groups of Galaxies 
({\bf Basel}: Gordon and Breach Science Publishers)
\reference{} Hudson, M. J., Lucey, J. R., Smith, R. J., \& Steel, J. 1997, \mnras, 291, 488
\reference{} J\o rgensen, I., Franx, M., \& Kj\ae rgaard, P. 1995a, \mnras, 273, 1097
\reference{} J\o rgensen, I., Franx, M., \& Kj\ae rgaard, P. 1995b, \mnras, 276, 1341
\reference{} J\o rgensen, I., Franx, M., \& Kj\ae rgaard, P. 1996, \mnras, 280, 167
\reference{} J\o rgensen, I., Franx, M., Hjorth, J., \& van Dokkum, P. G. 1999, \mnras, 308, 833
\reference{} Lynden-Bell, D., Faber, S. M., Burstein, D., Davies, R. L., Dressler, A., Terlevich, R. J., \& Wegner, G. 1988, \apj, 326, 19
\reference{} McElroy, D. B. 1995, \apjs, 100, 105
\reference{} Pahre, M. A., 1998, PhD Thesis, California Institute of Technology
\reference{} Pahre, M. A., Djorgovski, S., \& de Carvalho, R. R. 1998a, \aj, 116, 1591
\reference{} Pahre, M. A., Djorgovski, S., \& de Carvalho, R. R. 1998b, \aj, 116, 1606
\reference{} Rix, H.-W., \& White, S. D. M. 1992, \mnras, 254, 389
\reference{} Saglia, R. P., Bender, R., \& Dressler, A. 1993, A\&A, 279, 75
\reference{} Sargent, W. L. W., Schecter, P. L., Boksenberg, A., \& Shortridge, K. 1977, \apj, 212, 326
\reference{} Scodeggio, M., Giovanelli, R., \& Haynes, M. P. 1997, \aj, 113, 101 \reference{} Tonry, J. L., \& Davis, M. 1979, \aj, 84, 1511
\reference{} Whitmore, B. C., McElroy, D. B., \& Tonry, J. L. 1985, \apjs, 59, 1
\reference{} Zepf, S. E. 1991, PhD Thesis, Johns Hopkins University
\reference{} Zepf, S. E., \& Whitmore, B. C. 1993, \apj, 418, 72 (ZW93)

\end{references}
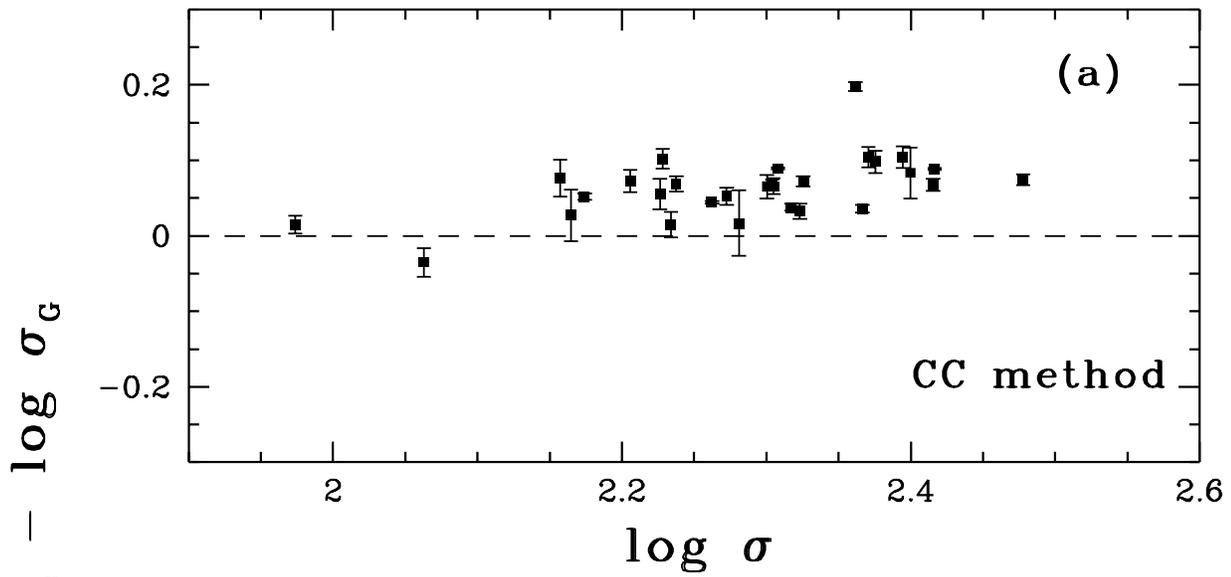
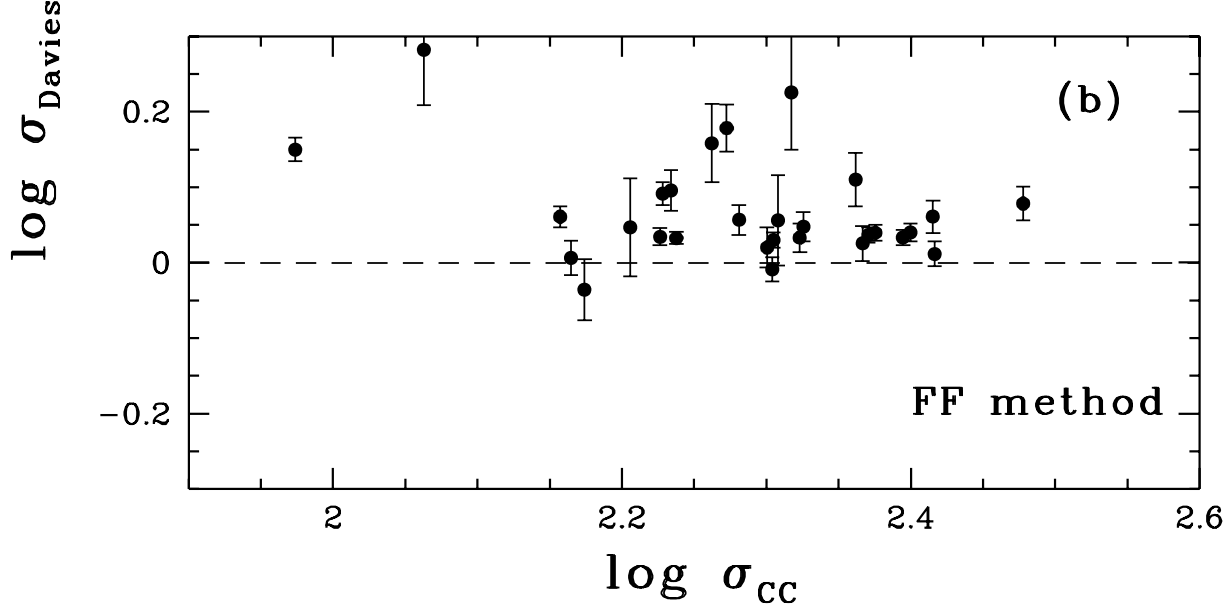
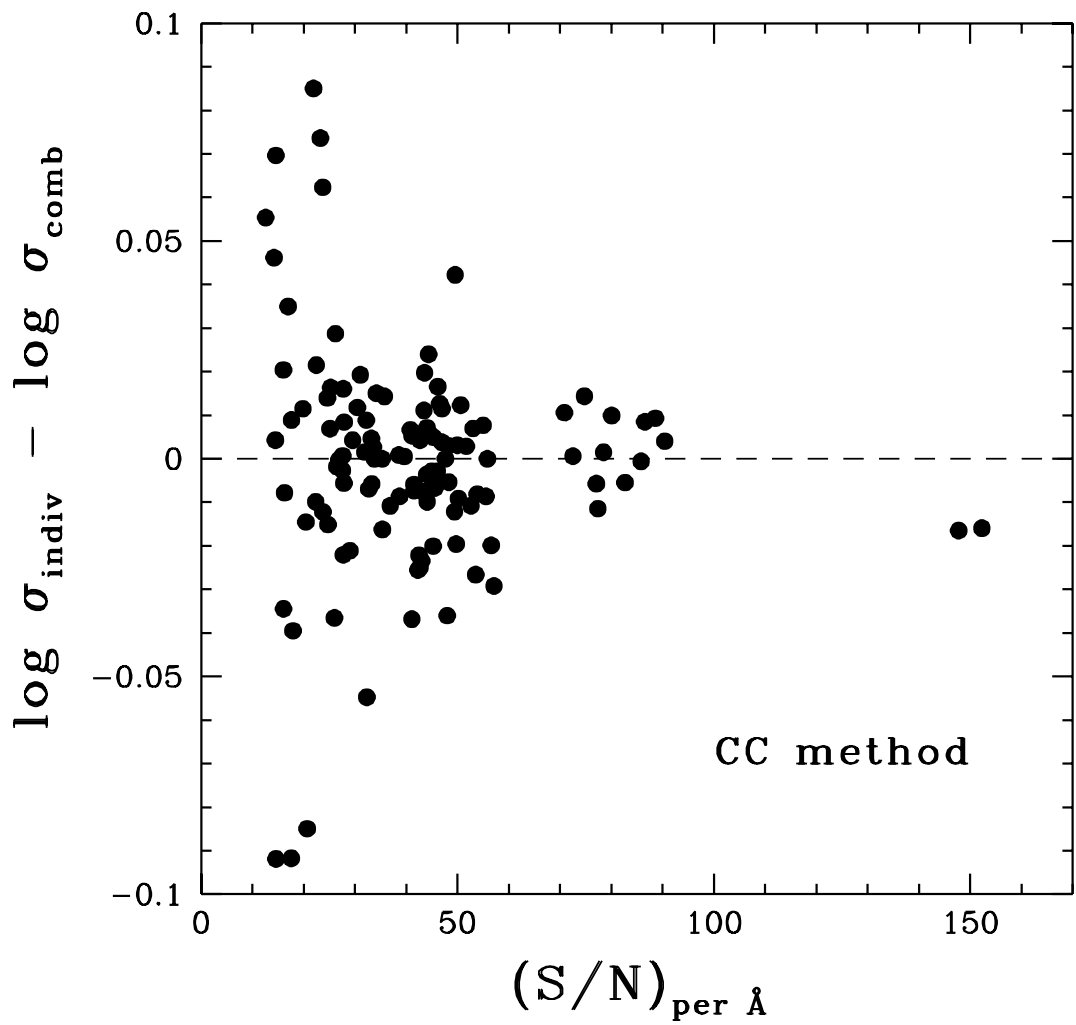
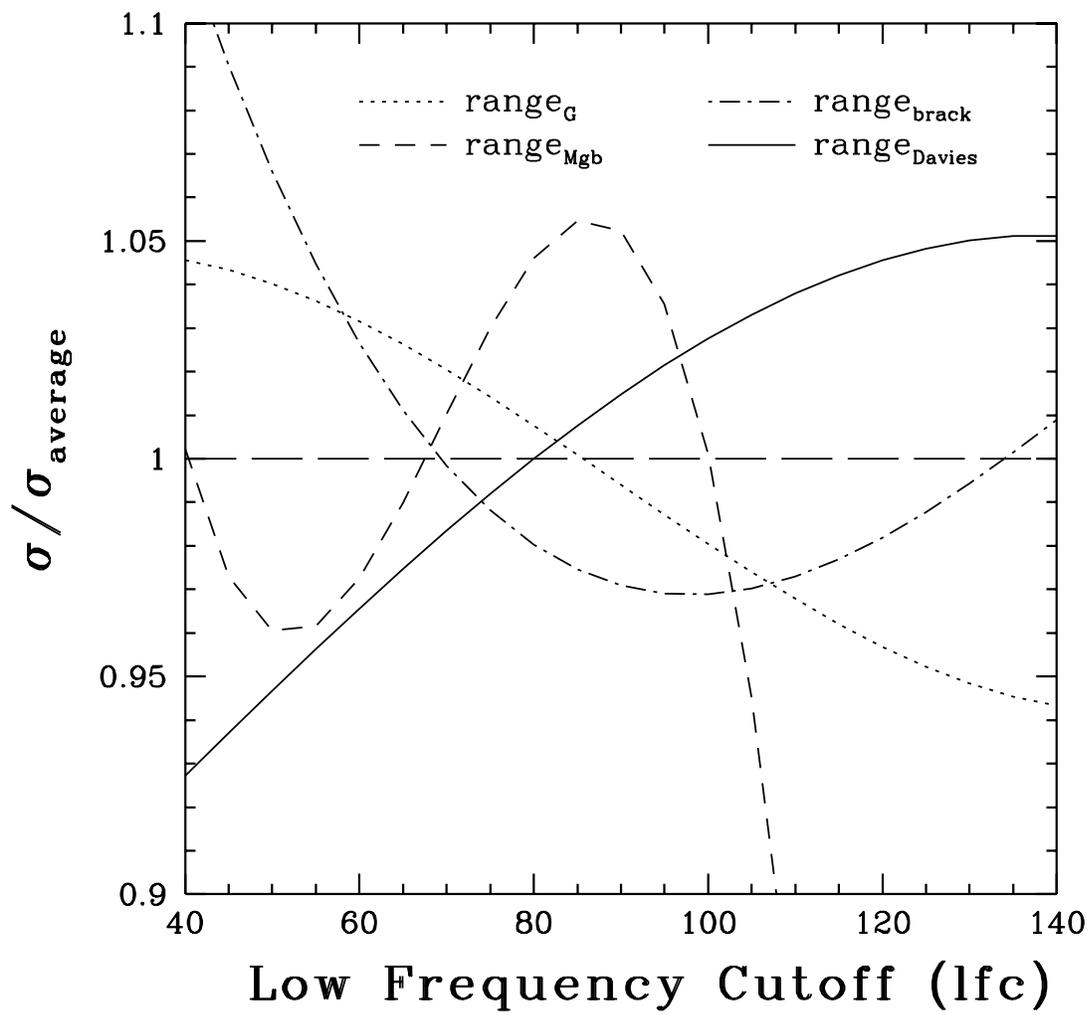
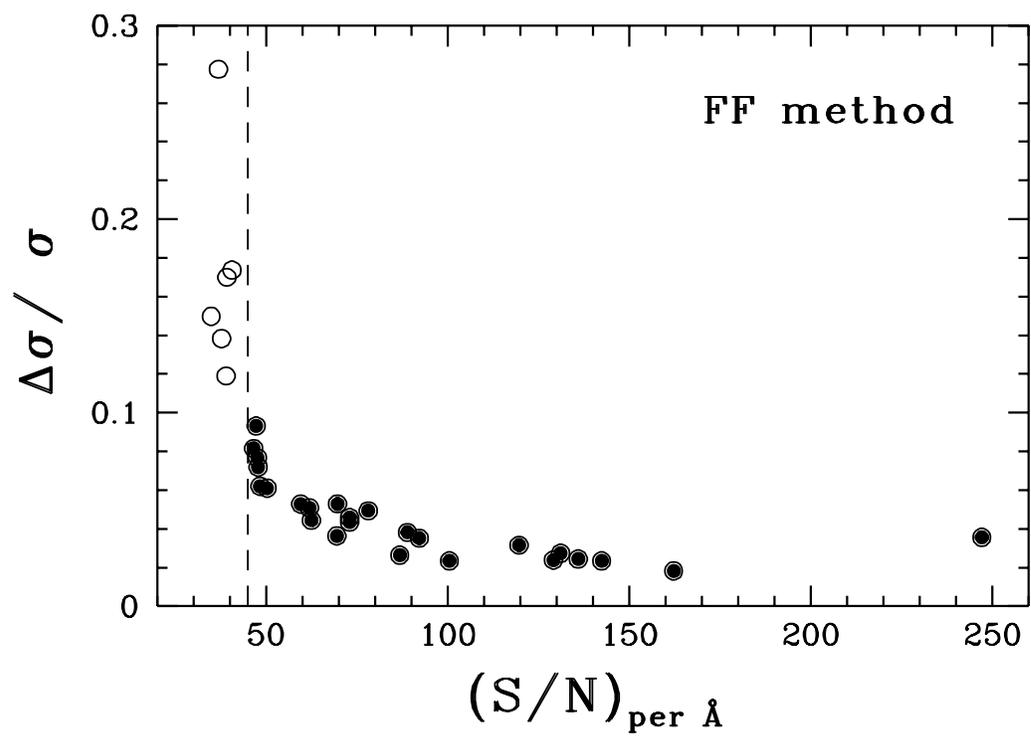
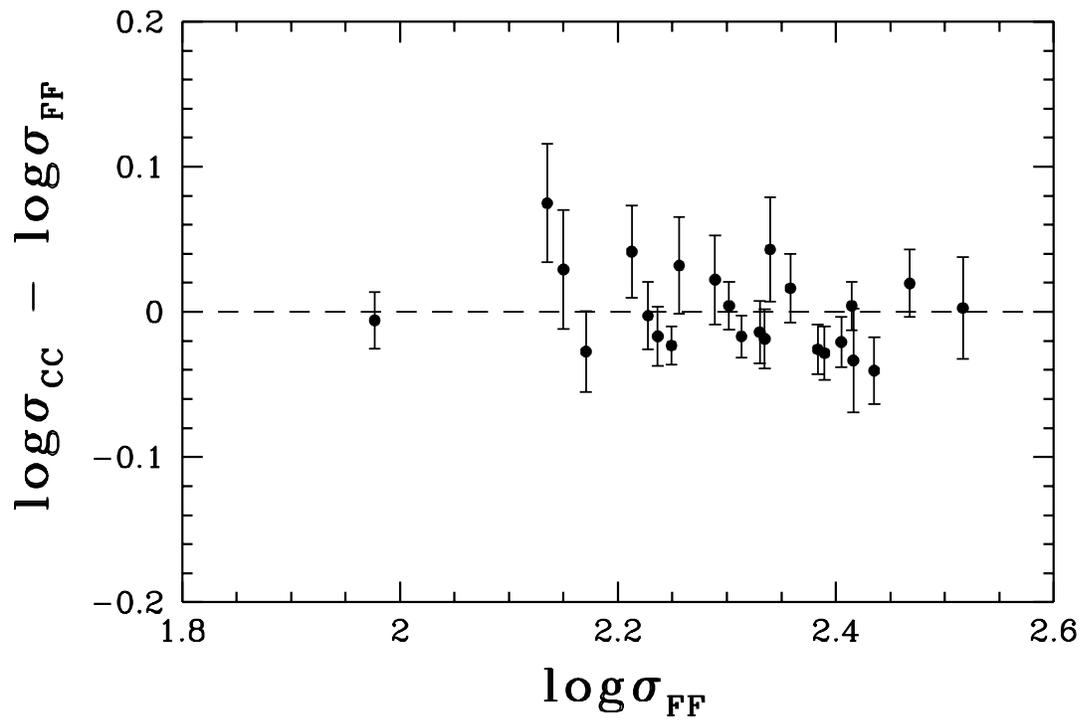
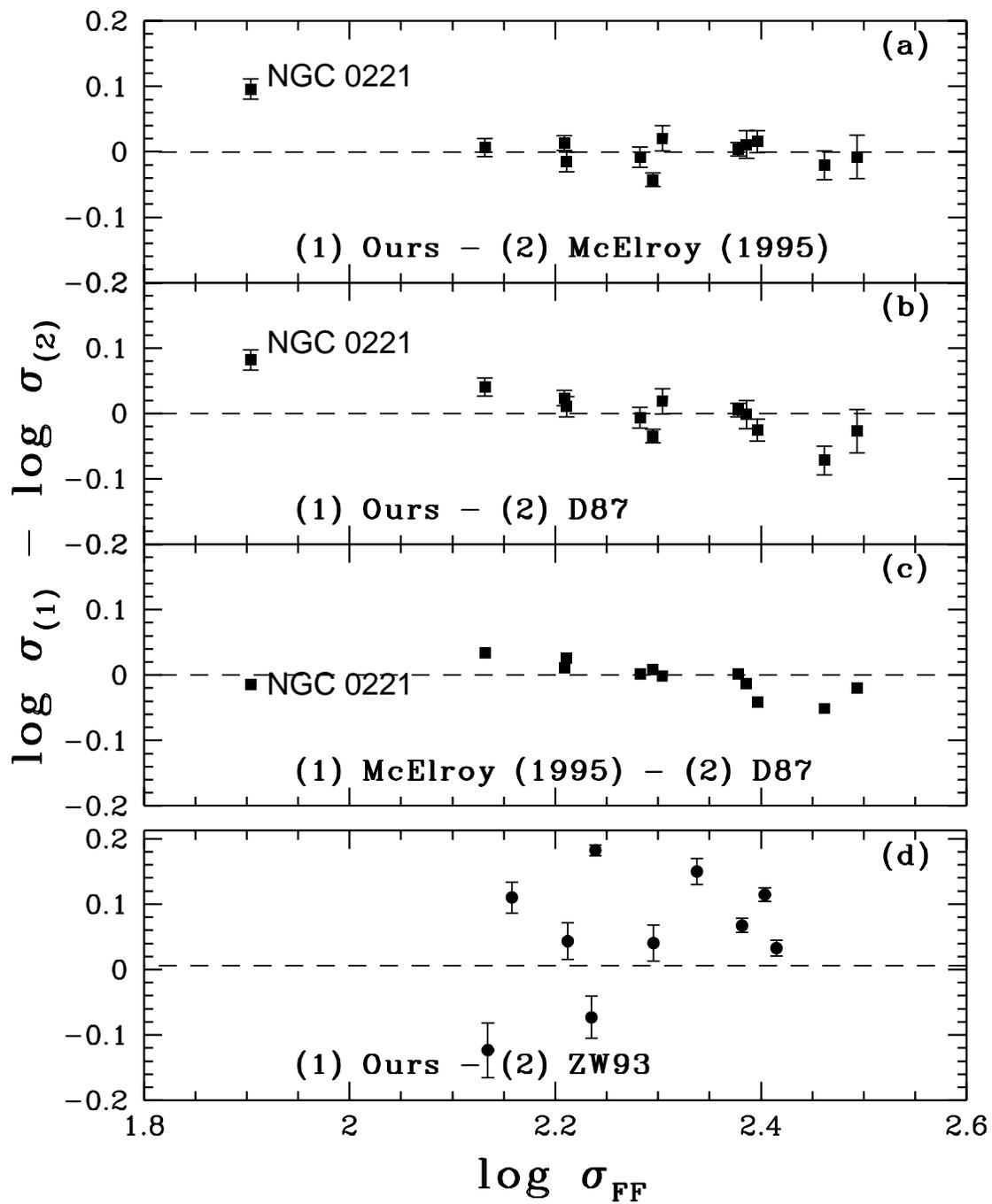
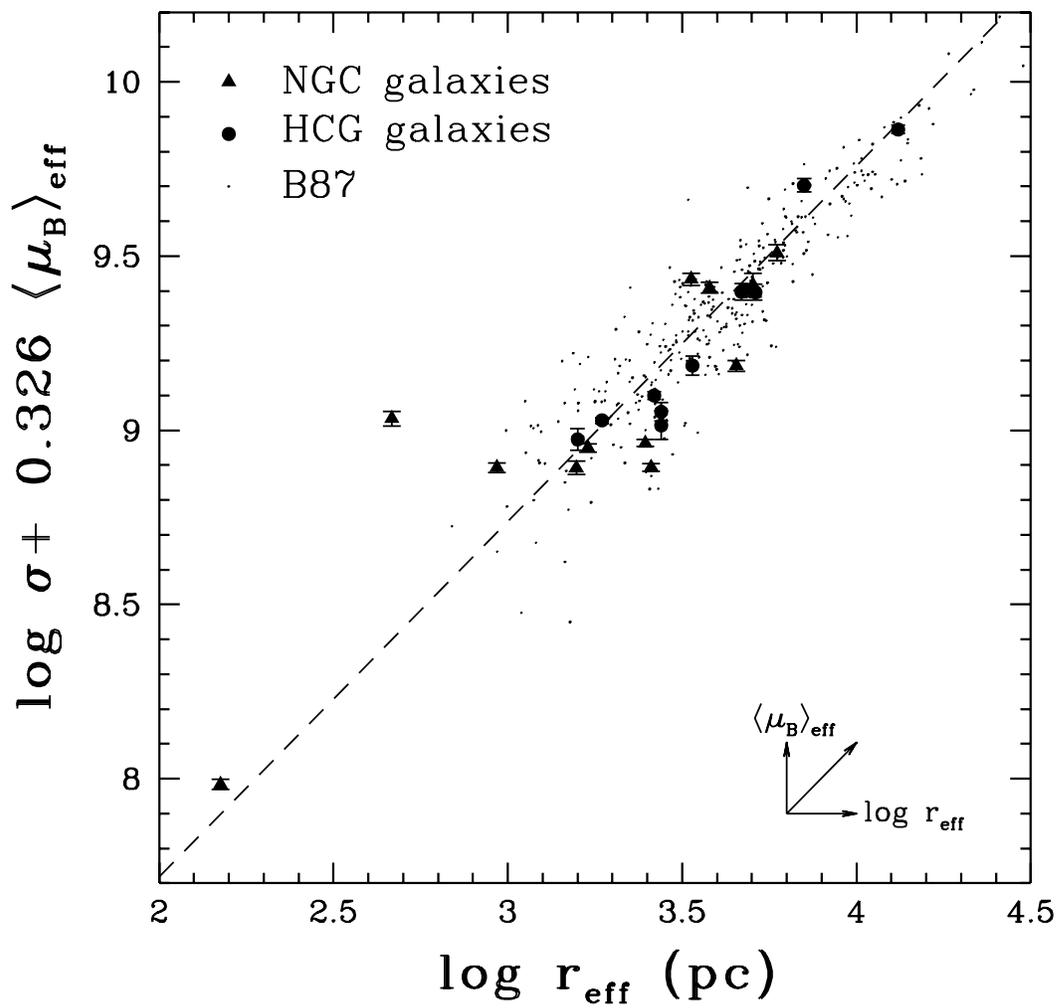
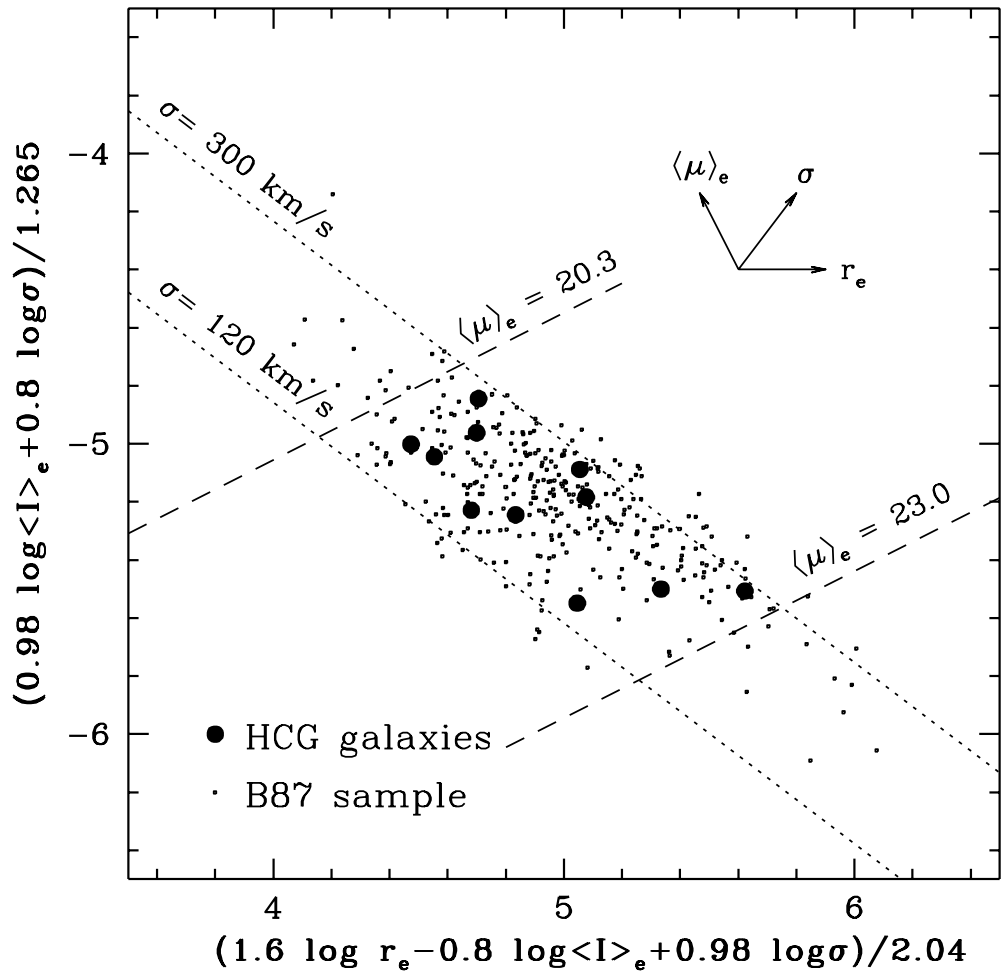
\end{document}